\documentclass[journal,twocolumn,twoside,10pt]{article}
\usepackage[left=1.7cm, right=1.7cm, top=2cm, bottom=2cm]{geometry}
\usepackage{graphicx}
\usepackage[utf8]{inputenc}
\usepackage{amsmath}
\usepackage{amsfonts}
\usepackage{amsthm}
\usepackage{hyperref}

\usepackage{xcolor}

\title{GACELA \\ \huge{A generative adversarial context encoder\\ for long audio inpainting}}
\author{Andr\'es Marafioti, Piotr Majdak, Nicki Holighaus, and Nathana\"el Perraudin
\renewcommand\footnotemark{}

\thanks{Manuscript received on May 2020;}%
\thanks{Andr\'es Marafioti, Piotr Majdak, and Nicki Holighaus are with the Acoustics Research Institute, Austrian Academy of Sciences, Wohllebengasse 12--14, 1040 Vienna, Austria.}
\thanks{Nathana\"el Perraudin is with the Swiss Data Science Center, ETH Z\"urich, Universit\"atstrasse 25, 8006 Z\"urich}
\thanks{Accompanying web page (sound examples, Python code, color figures):\newline
\texttt{\url{https://andimarafioti.github.io/GACELA/}}. \newline 
We specially thank Michael Mihocic for running the experiments at the Acoustics Research Institute's laboratory during the coronavirus pandemic as well as the subjects for their participation. \newline
This work has been supported by Austrian Science Fund (FWF) project MERLIN (Modern methods for the restoration of lost information in digital signals;I 3067-N30). We gratefully acknowledge the support of NVIDIA Corporation with the donation of the Titan X Pascal GPU used for this research.}}
\begin{document}

\maketitle
\begin{abstract}
We introduce GACELA, a generative adversarial network (GAN) designed to restore missing musical audio data with a duration ranging between hundreds of milliseconds to a few seconds, i.e., to perform long-gap audio inpainting. While previous work either addressed shorter gaps or relied on exemplars by copying available information from other signal parts, GACELA addresses the inpainting of long gaps in two aspects. First, it considers various time scales of audio information by relying on five parallel discriminators with increasing resolution of receptive fields. Second, it is conditioned not only on the available information surrounding the gap, i.e., the context, but also on the latent variable of the conditional GAN. This addresses the inherent multi-modality of audio inpainting at such long gaps and provides the option of user-defined inpainting. GACELA was tested in listening tests on music signals of varying complexity and gap durations ranging from 375~ms to 1500~ms. While our subjects were often able to detect the inpaintings, the severity of the artifacts decreased from unacceptable to mildly disturbing. GACELA represents a framework capable to integrate future improvements such as processing of more auditory-related features or more explicit musical features.  
\end{abstract}

\section{Introduction}

Audio signals frequently suffer from undesired localized corruptions. These corruptions can be a product of issues during the recording, packet-loss in transmission, or a faulty storage such as a scratched vinyl record. Although the corruptions may have different roots, the study of their removal can be unified as the restoration of localized lost information, usually called \emph{audio inpainting}~\cite{Adler2012}. This restoration has also been referred to in the literature as audio interpolation and extrapolation~\cite{kauppinen2001method,Etter1996}, or waveform substitution~\cite{Goodman1986}. For short corruptions affecting less than a few tens of milliseconds, the goal of audio inpainting algorithms has been to recover the lost information exactly~\cite{mokry2019introducing}. But as the corruptions affect longer periods of time, that goal becomes unrealistic. For long corruptions, audio inpainting algorithms attempt to reduce the damage by preventing audible artifacts and introducing new coherent information. The new information needs to be semantically compatible, a challenging task for for music, which often has a strict underlying structure with long dependencies. Previous work~\cite{Perraudin2016} tried to exploit this structure by repurposing information already available in the signal instead of generating new information. This approach has obivous downsides as it expects music to have repetitions and it usually modifies the length of the corruption. Others~\cite{marafioti2019context} have proposed methods that do generate new information, but aim at exact reconstruction and therefore fail to generalize for corruptions over tens of milliseconds.

In  this contribution, we propose a novel audio inpainting algorithm that generates new information and is specifically designed to address corruptions in the range between hundreds of milliseconds and seconds. In particular, we study the algorithm for the reconstruction of musical signals, i.e., a  mix of sounds from musical instruments organized in time. Further, we assume that areas of lost information are separated in time, such that the local information surrounding the gap, the \emph{context}, is reliable and can be exploited. The proposed algorithm relies on a generative adversarial network (GAN)~\cite{Goodfellow2014} conditioned on the encoded context information. We refer to the algorithm as a context encoder following~\cite{Pathak2016, marafioti2019context}. Our context encoder aims at generating content that matches the sound characteristics and respects the semantic cohesion of the available bordering information. In this contribution, we explain our design choices and we provide a thorough evaluation of our context encoder to determine factors with the largest potential for future improvement. Our software and trained models, complemented by instructive examples, is available at \url{https://andimarafioti.github.io/GACELA/}.

\subsection{Related audio-inpainting algorithms} 
The term "audio inpainting" was coined by Adler et al. to describe a large class of inverse problems in audio processing~\cite{Adler2012}. Their own work, however, mostly studied the restoration of gaps in audio signals. Generally, audio inpainting problems are concerned with audio represented as data in some \emph{feature} domain and assume that chunks of that data are corrupted yielding \emph{gaps} in the representation. 

The number and duration of the gaps as well as the type of corruption is manifold. For example, in declicking and declipping, corruptions may be frequent, but mostly confined to disconnected time-segments of few milliseconds duration or less. We refer to this as inpainting of \textit{short} gaps. On the other hand, gaps on a scale of hundreds of milliseconds or even seconds may happen, e.g., when reading partially damaged physical media, in live music recordings, when unwanted noise originating from the audience needs to be removed, or in audio transmission with a total loss of the connection between transmitter and receiver lasting for seconds. In such cases we attempt to inpaint \textit{long} gaps. 

For inpainting short gaps, various solutions have been proposed. \cite{Adler2012} proposed a framework based on orthogonal matching pursuit (OMP), which has inspired a considerable amount of research exploiting TF sparsity \cite{Adler2011, toumi:hal-01680669,kitic:hal-01159700,mokry2019introducing} or structured sparsity \cite{gaultier:hal-01540945,siedenburg:hal-01002998,Lieb2018}. As discussed in ~\cite{mokry2019introducing}, such methods do \emph{not} extend well to longer gaps, see also \cite{mokry2020audio} for a recent study of sparsity-based audio inpainting. Other methods for  short gap inpainting and relying on TF representations rely, e.g., on a regression model~\cite{wolfe2005interpolation}, or nonnegative matrix and tensor factorization~\cite{le2011computational,smaragdis2011missing,csimcsekli2012score}. More recently, a powerful framework has been proposed for various audio inverse problems~\cite{Bilen2018} including time-domain audio inpainting, source separation~\cite{bilen2015joint}, and declipping~\cite{bilen2015audio} even in a multichannel scenario~\cite{ozerov2016multichannel}. 

Interpolation algorithms based on linear prediction coding (LPC) ~\cite{Tremain1982} are flexible enough to cover various gap lengths, but pose strong assumptions on stationarity of the distorted signal~\cite{janssen1986adaptive,Etter1996,kauppinen2002audio}. Nonetheless, they outperform the aforementioned short gap methods on gap durations above $50$~ms~\cite{mokry2019introducing}. Despite a recent contribution by the authors proposing a neural context encoder to perform inpainting of medium duration gaps \cite{marafioti2019context,marafioti2019audio}, LPC is still, in our opinion, amongst the most promising methods for inpainting medium duration gaps.

On the other hand, for inpainting long gaps, i.e., gaps exceeding several hundred milliseconds, recent methods leverage repetition and determine the most promising reliable segment from uncorrupted portions of the input signal. Restoration is then achieved by inserting the determined segment into the gaps. These methods do not claim to restore the missing gap perfectly, they aim at \emph{plausibility}. For example, exemplar-based inpainting was proposed based on a graph encoding spectro-temporal similarities within an audio signal~\cite{Perraudin2016}. Other examples of long gap audio inpainting by means of exemplars include \cite{Bahat2015,manilow2017leveraging,martin2011exemplar,maher1994method,lukin2008parametric}. Not aiming for \emph{accurate recovery} of the missing information, but instead for a plausible solution, numerical criteria are ill-suited for assessing the success of long gap inpainting, which usually requires extensive perceptual evaluation. 

Starting around 2019, several groups of researchers have attempted to tackle the audio inpainting problem using deep neural networks and TF representations. In \cite{zhou2019vision}, spectrogram inpainting from combined audio and video information is proposed, while \cite{kegler2019deep} considers inpainting of time-frequency masked speech data. The context encoder presented in \cite{marafioti2019context,marafioti2019audio} is specifically targeted to medium duration gaps. The pre-print \cite{chang2019deep} seems to be the first to evaluate the inpainting of slightly longer gaps (100 to 400~ms) with neural networks, comparing several models originally proposed for general conditioned audio synthesis with the graph-based method \cite{Perraudin2016}, as well as the authors' own proposal for a deep neural system for audio inpainting. Generally speaking, deep audio inpainting seems to be a particularly tough instance of conditioned deep neural audio synthesis, since the conditioning only contains indirect information about the content to be generated, which nonetheless needs to be seamlessly inserted into the existing reliable audio.

\subsection{Related deep-learning techniques / audio synthesis with ML}

There have been many attempts to synthesize audio using neural networks. However, neural audio synthesis remains a particularly challenging task because of the presence of complex structures with dependencies on various temporal scales. Neural audio synthesizers are often conditioned to reduce the dependencies on larger temporal scales~\cite{Engel2017, Dieleman2018}, but even then the networks that finally synthesize the signal are fairly sophisticated~\cite{Wavenet2016, Mehri2017, prenger2019waveglow, MelGAN-2019, binkowski2019high, kalchbrenner18a}. This can be partially explained by these networks modeling audio as a time representation with a high temporal resolution; audio time signals usually have at least 16,000 samples per second. In contrast, when modeling audio as a time-frequency (TF) representation, the temporal resolution is a parameter of the model. In fact, TF representations of audio are widely applied to neural networks, e.g., for solving discriminative tasks, in which they outperform networks directly trained on the waveform \cite{dieleman2014end, Pons2017, Abbasi2019ApplyingCN}. TF representations are also commonly chosen to condition neural synthesizers~\cite{saito2018text, jin2018fftnet}, e.g., Tacotron 2~\cite{Shen2018} relies on non-invertible mel-frequency spectrograms and Timbretron~\cite{huang2018timbretron} relies on the constant-Q transform. In those cases, the generation of a time-domain signal from the TF coefficients is then achieved by training a conditional neural synthesizer to act as a vocoder. Despite recent improvements in neural synthesizers modeling audio in the TF domain~\cite{engel2019gansynth, marafioti2019adversarial, vasquez2019melnet}, the state-of-the-art neural synthesizers still model audio in the time domain.

We can obtain valuable insights on the design of a neural synthesizer for audio inpainting from music synthesizers. Modeling music has proven particularly difficult due to a wide range of timescales in dependencies from pitch and timbre (short-term), through rhythm (medium-term) to song structure (long-term)~\cite{hawthorne2018enabling, Dieleman2018}. The long-range dependencies can be addressed by synthesizing music in multiple steps. Different features have been proposed as intermediate representations~\cite{Herremans2017, Blaauw2017, Chowdhury2019}, with a common symbolic one being MIDI~\cite{Boulanger2012ModelingTD, hawthorne2018enabling, Manzelli2018}. Conditioning neural synthesizers with neurally generated MIDI has many advantages: 1) it is analogous to the discrete structure embedded in music's generative process, in the words of \cite{hawthorne2018enabling}: ''a composer creates songs, sections, and notes, and a performer realizes those notes with discrete events on their instrument, creating sound``. 2) MIDI is easy to interpret and modify. Users can interact with MIDI pieces generated from a network before the neural synthesizer plays them. On the other hand, MIDI's major drawback is that in order to learn from it, one needs  large-scale annotated datasets. For piano music, \cite{hawthorne2018enabling} addressed this by creating such a dataset, but for general music we have not found a suitable dataset.

\section{The inpainting system: GACELA} 

\begin{figure*}[!th]
\begin{center}
	\includegraphics[width=0.8\textwidth]{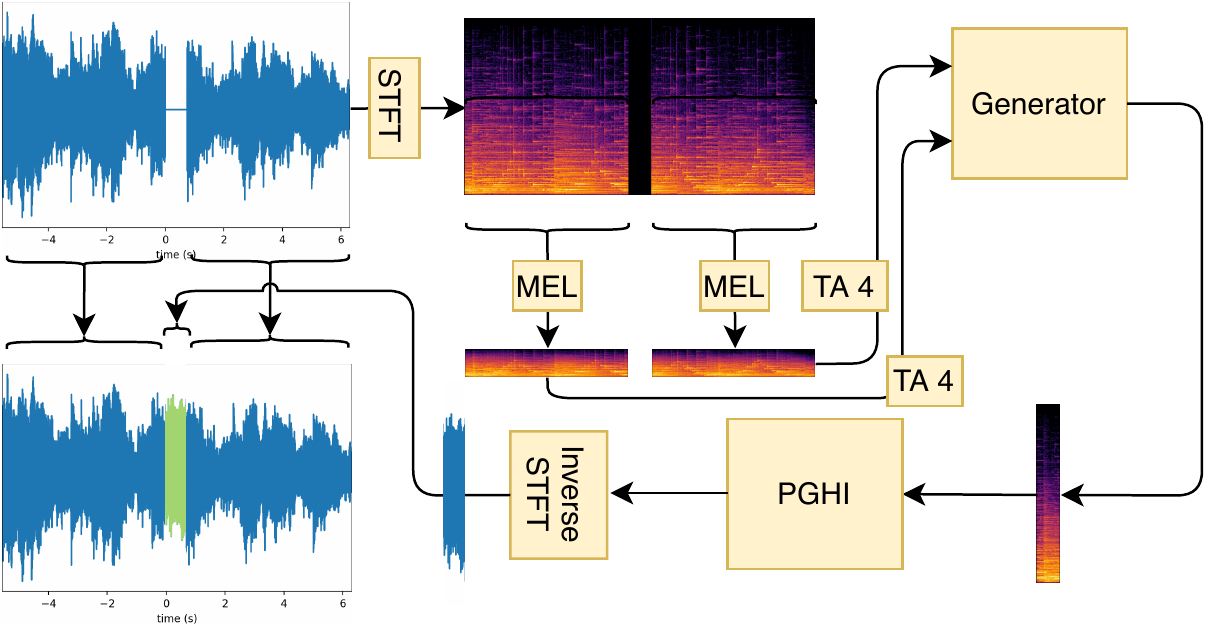}
	\caption{Overview of the end-to-end audio generation system. The discriminators are only used for training the generator. The phase construction extends the phase from the real context into the generated spectrogram.}
	\label{fig:overview-system}
\end{center}
\end{figure*}

Our generative adversarial context encoder (GACELA) targets music inpainting in long gaps, i.e., in the range between hundreds of milliseconds and seconds. In this range, there are usually multiple plausible solutions for music inpainting and we consider the task as multimodal. For example, on a gap where originally a single chord was played, there could be several other chords that fill in the gap while still sounding plausible. For each chord there are even several variations: different intensities or onsets for each note. The multi-modality present at this range needs to be taken into account to model the task. Considering that a standard regression loss models a unique solution, it would lead to an average of the possible solutions and it is a bad fit for the task at this range. To solve this challenge, we model the task with a GAN, which is able to model the distribution of possible gap replacements instead of producing a single candidate. GANs rely on two competing neural networks trained simultaneously in a two-player min-max game: The generator produces new data from samples of a random variable; The discriminator attempts to distinguish between these generated and real data. During the training, the generator's objective is to fool the discriminator, while the discriminator attempts to learn to better classify real and generated (fake) data.

An overview of our end-to-end audio generation is presented in Fig.~\ref{fig:overview-system}. As in \cite{marafioti2019context}, we consider the audio signal $s$ consisting of the gap $s_g$ and the context signals before and after the gap, $s_b$ and $s_a$, respectively. The signal $s$ is transformed into mel-scale time-frequency spectrograms (mel spectrograms) and unreliable time frames, i.e., those that have nonnegligible overlap with the gap, are discarded. The remaining mel coefficients form the preceding context $S_b$ and succeeding context $S_a$. After further dimensionality reduction, the contexts serve to condition the generator. The output of the generator is a log-magnitude STFT $S_g$, from which an audio signal is synthesized using established methods for phaseless reconstruction. 

The proposed adversarial context encoder is comprised of a generator network and five discriminator networks, which consider the (generated or real) audio content in the gap region and its context at different scales. Each discriminator receives the generated (or real) gap data, as well as different amounts of context, encoded either as log-magnitude short-time Fourier coefficients or mel spectrograms, depending on the scale of the considered context. The generator, a context encoder conceptually split into two identical border encoder networks and one decoder network, is \emph{conditioned} on the real border data encoded as mel spectrograms. Such conditioning is commonly achieved in GANs by supplying auxiliary data to both the generator and the discriminator, further specifying the generator's task. To date, several formulations of conditional architectures have been proposed~\cite{Miyato2018}. In this contribution, we opt to condition solely on the close-range context, i.e., few seconds of TF audio data preceding and succeeding the gap. Converting the time-domain audio into a log-magnitude time-frequency representation partially solves the \emph{problem of scales}, since a large number of audio samples are represented by a small number of time frames in the time-frequency representation. Hence, all audio data is transformed into, and represented by, log-magnitude short-time Fourier spectrograms and sometimes further processed into mel spectrograms~\cite{rabiner2011theory}. The latter is the de-facto standard for a perceptually motivated dimensionality-reduced time-frequency representation and is well-suited as a basic encoding for larger scale conditioning data, when reduced precision is sufficient or even desired. The former on the other hand provides a redundant, highly detailed and interpretable representation of audio from which the source signal can be reconstructed in excellent quality by means of recent algorithms for phaseless reconstruction \cite{pruuvsa2017noniterative,ltfatnote048,ltfatnote021,leroux10}. 




The software was implemented in PyTorch~\cite{pytorch} and is publicly available, as well as every trained model here discussed.\footnote{\url{www.github.com/andimarafioti/GACELA}} The audio processing blocks described in the next subsection are computed using the 
Tifresi package~\cite{tifresi}. 
For computing the STFT and mel representations, Tifresi depends on LTFATpy~\cite{ltfatweb} and librosa~\cite{librosa}, respectively. 

\subsection{Processing stages}

In addition to the network architecture, the proposed training- and generation-time pipelines require some simple, fixed signal processing blocks to transform the data at various points in the processing chain. 

\begin{itemize}
 \item \textbf{STFT:} Computes the log-magnitude STFT spectrogram of the input audio waveform. Optionally, the STFT phase can be stored for later use. In the provided implementation, all STFTs are computed using truncated Gaussian windows with a hop size of $a= 256$ and $M = 1024$ frequency channels, following the guidelines proposed in~\cite{marafioti2019adversarial}, leading to a representation with redundancy $M/a = 4$.
 \item \textbf{ISTFT:} Given a log-magnitude STFT and matching STFT phase, the ISTFT block performs STFT inversion. Inversion is matched to the STFT block in the following sense: When log-magnitude and phase input equal the output of STFT, the waveform output of ISTFT equals the waveform input of STFT. See \cite{groechenig2001foundations, christensen2002introduction, feichtinger1997gabor} for more information on how to invert redundant STFTs.
 \item \textbf{PGHI:} Constructs a candidate phase for a given log-magnitude STFT spectrogram by means of phase gradient heap integration~\cite{pruuvsa2017noniterative}. The output phase can be combined with the input spectrogram for use with ISTFT.
 \item \textbf{MEL:} Computes a mel-scale spectrogram from a given log-magnitude STFT spectrogram. In the provided implementation, all mel spectrograms are set to have $80$ filters.
 \item \textbf{Time-Averaging (TA X):} Reduces the time dimension of a log-magnitude STFT spectrogram or MEL spectrogram by a factor of $X$, where $X$ is a positive integer. Dimensionality reduction is achieved by averaging every $X$ successive time frames of the input spectrogram.
\end{itemize}

At generation time, the audio processing blocks are used to preprocess the generator input and during audio synthesis from the generator output as shown in Figures \ref{fig:overview-system} and \ref{fig:generator}. During training, the discriminator input is preprocessed as well, as shown in Figure \ref{fig:discriminators-overview}. 

We use PGHI~\cite{pruuvsa2017noniterative} over competing phaseless reconstruction algorithms such as Griffin-Lim~\cite{griffin1984signal} or LeRoux's weighted least squares~\cite{leroux10} for various reasons: PGHI is non-iterative, highly efficient and often outperforms other algorithms in terms of perceptual reconstruction quality, sometimes even significantly~\cite{pruuvsa2017noniterative,ltfatnote043,ltfatnote048}.

\subsection{The adversarial context encoder}

%

\begin{figure*}[!th]
\begin{center}
	\includegraphics[width=0.8\textwidth]{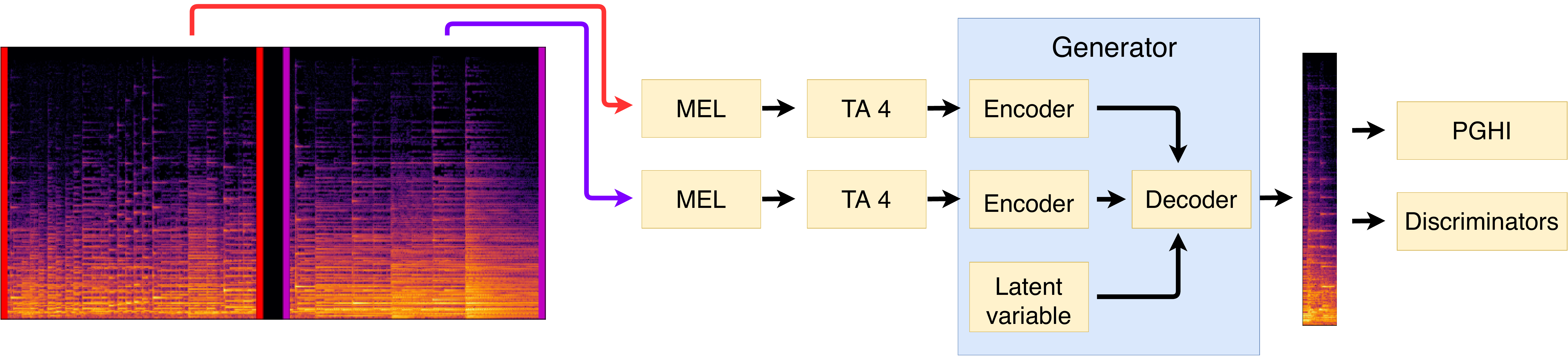}
	\caption{Overview of the generation process. At training time, the generator output is forwarded to the discriminators. During generation, an audio waveform is generated by processing the output with the PGHI and ISTFT blocks.}
	\label{fig:generator}
\end{center}
\end{figure*}

\textbf{Generator:} The overall structure of the generator is shown in Fig. \ref{fig:generator}.
The preceding and succeeding contexts are each provided to their associated encoder network after passing through the preprocessing chain, consisting of an STFT block, followed by a MEL and TA 4 block. Both encoders share the same architecture: $4$ convolutional layers with stride $2$ and ReLU activations. Parameters are not shared between the encoders.
Both encoder outputs and a realization of the latent variable are concatenated and passed to the decoder. The latent variable is drawn from a $128$-dimensional uniform distribution. The decoder itself is comprised of a fully-connected layer with output size 8192, followed by $4$ transposed convolution layers with stride $2$, further $2$ convolutional residual layers and a final transposed convolution layer. The decoder output has exactly the shape of the gap in the original log-magnitude STFT spectrogram and is interpreted as log-magnitude STFT coefficients for the purpose of synthesis and propagation through the discriminators. For more details about the generator, refer to the accompanying code implementation.

\begin{figure}[!h]
\begin{center}
	\includegraphics[width=0.8\columnwidth]{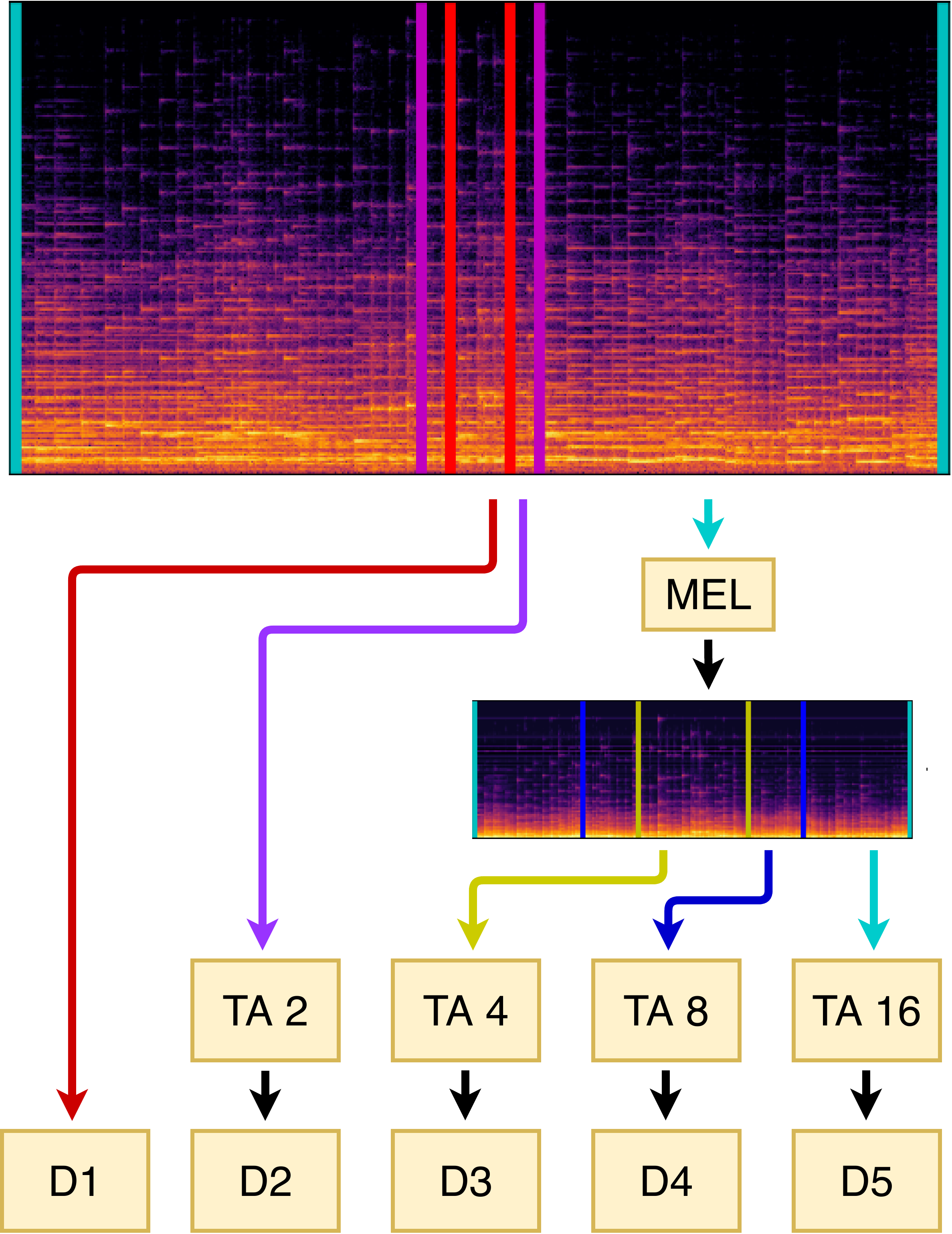}
	\caption{Overview of the multi-scale architecture from the discriminator. D1-D5 represent the individual discriminators. The different receptive fields of the discriminators are marked in colors. The center, marked with red lines, is the input of the first discriminator and contains only the generated or real gap.}
	\label{fig:discriminators-overview}
\end{center}
\end{figure}

\textbf{Discriminators:} 
We adapt the multi-scale architecture from \cite{wang2018high} to TF representations, with five discriminators operating on five different time scales and two different frequency representations. In the audio domain, multiple discriminators operating on different scales have successfully been used as well \cite{MelGAN-2019}, where the authors directly process time-domain audio. In the proposed architecture, all discriminators are supplied with time-frequency spectrograms and between successive discriminators, the receptive field is increased by a factor of two, see Fig. \ref{fig:discriminators-overview}. Every discriminator has five convolutional layers with stride $2$.
While the first two discriminators, with smaller receptive field, process log-magnitude STFT spectrograms directly, further discriminators process mel coefficients. Since the increased receptive field is achieved by time-averaging, the number of input time frames is equal for all discriminators. On the other hand, the input mel spectrograms supplied to discriminators 3 through 5 posses a reduced number of frequency channels, such that these discriminators were allocated $4$ times less channels in every convolutional layer. For more details about the discriminators, refer to the accompanying code implementation.

\section{Evaluation methods}

The main objective of the evaluation is to determine to which extent our system can restore localized corruptions in different types of musical signals. We want to determine 1) the effect of the complexity level of the musical signal on the inpainting performance and 2) the  effect of the gap length on the inpainting performance. To address 1), we built five different datasets of musical signals with increasing complexity from two types of audio signals: audio synthesized from MIDI files and recorded from physical instruments. For 2), the system was trained with an inserted gap size of either 375ms, 750~ms or 1,5~s. For these two tests, we evaluated the inpainting quality by means of listening tests.

The second objective of the evaluation is to better understand how the system performs the inpainting to facilitate improving it in the future and to gain insight for the development of other similar systems. Since our system relies on PGHI for phase reconstruction, we evaluated the impact of the PGHI on the audio inpainting. Further, we investigate the influence of the latent variable on the generator and the processing of context data by the encoder. Finally, we compare our method to the one presented in \cite{Perraudin2016}.

\subsection{Complexity levels}

We expect the success of our method to be correlated to the complexity of the music it is trained on. We tested this hypothesis with listening tests. To do so, we trained the method on different datasets with increasing complexity. The first two datasets were synthesized from MIDI data using pretty\_midi~\cite{prettymidi}, specifically it's fluidsynth API. We generated just one instrument and set that instrument to the piano program 1 so that the whole dataset would have the same sound complexity and to reduce the variability in the datasets. In total, we trained networks with five complexity levels, out of which four surpassed an initial informal evaluation, such that they were considered for the listening tests. Therefore, we tested 4 complexity levels, with 3 conditions per case, and 12 songs, giving us a total of 144 stimuli per block for this test.

\textbf{1) Simple midi. }
The simplest case we handled was `hand-written' MIDI data. Here, the MIDI annotations have little variation since they are written down by humans on a quantized structure. For this case, we used the Lakh MIDI dataset~\cite{raffelthesis}, a collection of 176,581 unique MIDI files, 45,129 of which have been matched and aligned to entries in the Million Song Dataset~\cite{millionsong}. The Lakh MIDI dataset was generated with the goal of facilitating large-scale music information retrieval, both symbolic (using the MIDI files alone) and audio content-based (using information extracted from the MIDI files as annotations for the matched audio files). 

\textbf{2) Midi recorded from human performances. }
For the second complexity level, we used MIDI data that was extracted from performances on a piano. Here, the added complexity is the lack of a strict musical structure such as the precise tempo present from level 1). For this case, we used the Maestro dataset~\cite{hawthorne2018enabling}, a dataset containing over 200 hours of paired audio and MIDI recordings from ten years of International Piano-e-Competition. In this competitions, virtuoso pianists perform on Yamaha Disklaviers which, in addition to being concert-quality acoustic grand pianos, utilize an integrated high-precision MIDI capture and playback system. The MIDI data includes key strike velocities and sustain/sostenuto/una corda pedal positions. 
The repertoire is mostly classical, including composers from the 17th to early 20th century.

\textbf{3) Audio recordings of piano performances. } 
For the third complexity level, we used real recorded performances of grand pianos. These are the same pieces from the second complexity level. This level adds the sound complexity of a real instrument compared to a simple midi synthesized sound. 

\textbf{4) Free music. }
The fourth level of complexity is the last one we used for the listening tests. For this, we wanted to test the system on a broader scenario including a more general definition of music. On this level, the added complexity is the interaction between several real instruments. To remove some variation from the dataset, we trained the network on a single genre at the time, in this case either rock or electronic music (for the listening test we only used rock samples). For this complexity level, we used the free music archive dataset (FMA,~\cite{fma2017}), particularly, a subset we generated by segmenting the `small' dataset by genre. FMA is an open and easily accessible dataset, usually used for evaluating tasks in musical information retrieval. The small version of FMA is comprised of 8,000 $30$-s segments of songs with eight balanced genres sampled at $44.1$~kHz.

\subsection{Gap durations}

We expect the success of our method to be correlated to the length of the gap. To test this hypothesis, we trained different networks on different gap lengths and evaluated them with listening tests. We kept the network structure as fixed as possible, such that Fig. \ref{fig:discriminators-overview} still applies to every network trained for this experiment. The selected gap lengths were either 372 ms, 743 ms or 1486 ms. Since we expect the effect of the gap length to be independent from the effect of the complexity of the music, we trained all networks on the third complexity level: real piano recordings. To evaluate the gap durations, we included two additional conditions to the listening tests\footnote{For this test, we did not need to include an additional 60 stimuli since the 743~ms gaps, the clicks, and the real signals were already considered on the complexity level.}. 

\subsection{Listening tests}
We performed listening tests to determine the effects of the complexity level and the gap length on the inpainting performance. 

\textbf{Subjects.} Candidates completed a self-assessment questionnaire about their music listening habits. For the evaluation, only candidates who listened to at least 4 hours of music per week were considered. In total, 8 subjects were selected for the test. They were paid on an hourly basis. Before the experiment, the subject was informed about the purpose and procedure of the experiment and five exemplary files were presented: 1) a sound with a gap, 2) the same sound with a click, 3) the same sound with a poor reconstruction, 4) the same sound with a good but detectable reconstruction, and 5) the original sound. Any questions with respect to the procedure were clarified.

\textbf{Task.} The task was similar to that from \cite{Perraudin2016}. In each trial, the subject listened to a sound stimulus and
was asked to pay attention to a potential artifact. A slider scrolled horizontally while the sample was played indicating the current position within a stimulus. The subject was asked to tag the artifact’s position by aligning a second slider with the beginning of the perceived artifact. Then, while listening again to the same stimulus, the subject was asked to confirm (and re-align if required) the slider position and answer the question "How poor was it" The possible answers were: (0) no issue ("Kein Fehler"), (1) not disturbing ("Nicht störend"), (2) mildly disturbing  ("Leicht störend"), and (3) not acceptable ("Nicht akzeptabel"). Then, the subject continued with the next trial by tapping the "next" button.

\textbf{Conditions.} Three conditions were tested: inpainted, clicked and reference (original). For the inpainted condition, the song was corrupted at a random place with a gap and then reconstructed with our method. The reconstructed song was cropped 2 to 4 seconds (randomly varying) before and after the gap resulting in samples of 4.4 to 9.5-s duration. For the reference condition, the same cropped segment was used. The reference condition did not contain any artifact and was used to estimate the sensitivity of a subject. For the click condition, a click was superimposed to the cropped segment at the position where the random gap started. The artifact in this condition was used as a reference artifact and was clearly audible. 

Across our datasets, the differences between songs are larger that in a single song, so we do not test the same song more than once. Instead, for each test 12 songs were used. The combination of complexity levels and gap lengths described in the remainder of this section resulted in a block of 168 stimuli. All stimuli were normalized in level. Within the block, the order of the stimuli and conditions was random. Each subject was tested with two blocks, resulting in 336 trials per subject in total. Subjects were allowed to take a break at any time, with two planned breaks per block. For each subject, the test lasted approximately three hours.

\subsection{Objective difference grade (ODG)}

In order to evaluate the influence of the phase reconstruction algorithm, PGHI, we computed the objective difference grade (ODG, \cite{recommendation20011387}), which corresponds to the subjective difference grade used in human-based audio tests, derived from the perceptual evaluation of audio quality. ODG ranges from $0$ to $-4$ with the interpretation shown in Tab.~\ref{tab:ODG}; it was computed using the implementation provided with ~\cite{kabal2002examination}. 
In our evaluation, the ODG was calculated on signals with the phase in the gap discarded and reconstructed using PGHI.

\begin{table}[!th]
	\centering
	\begin{tabular}{ll}
		\hline
ODG & Impairment
\\ \hline
0 & Imperceptible		 \\ 
-1 & Perceptible, but not annoying \\
-2 & Slightly annoying \\
-3 & Annoying\\
-4 & Very annoying \\
 \hline \\
	\end{tabular}
	\caption{Interpretation of ODG.}
	\label{tab:ODG}
	\vspace{-1em}
\end{table}

\section{Results}

\subsection{Impact of the phase reconstruction}

In order to assess the impact of the phase reconstruction on the inpainting quality, we evaluated the ODG of the real signals against signals which consisted of unaltered magnitude coefficients and PGHI applied only to recover the phase on the gap. This way, we were able to estimate the impact the PGHI will have for an output of the network that does not present problems at a STFT level. Additionally, since we apply ODG to the full signals used on the listening tests, we also compute the ODG for the click signals, to corroborate that ODG is sensitive to localized distortions.

The mean ODG obtained for 64 songs in the datasets used for the listening tests for both PGHI applied to the phase coefficients in the gap and a click applied at the beginning of the gap is presented in Table \ref{tab:pghi_odg}. From this, we can see that on the two MIDI datasets and the maestro recordings, the effect of PGHI would be very hard for listeners to detect, and even then it would not be annoying. On the other side, the click would always be easy to detect and annoying. For the most complex dataset, i.e., free music, the influence of PGHI was between imperceptible and perceptible but not annoying, and the influence of the click was perceptible and between not annoying and slightly annoying. This indicates that for the considered datasets, the effect of PGHI on the overall quality of the results will be small.

\begin{table}[]
    \centering
    \begin{tabular}{l|c|c}
          \hline
         Complexity level & ODG (PGHI)  & ODG (Click)   \\
          \hline
        1) Simple MIDI & -0.109 & -3.286 \\
        2) Recorded MIDI & -0.125 & -3.091  \\
        3) Recorded piano  & -0.231 & -3.503 \\
        4) Free music  & -0.618 & -1.732  \\ \hline 
    \end{tabular}
    \vspace{1em}
    \caption{PGHI's mean ODG accross 64 songs for different datasets}
    \label{tab:pghi_odg}
\end{table}

\subsection{Effect of the latent variable}

We condition the generator not only on the encoded context, but also on the latent variable, a random variable drawn from a uniform distribution. We expect different realizations of the random variable to output different solutions for the task. However, this might not be the case: It has been reported that GANs with strong conditioning information do not rely heavily on the additional noise input distribution \cite{MathieuCL15, isola2017image, MelGAN-2019}. In order to evaluate whether the generator output changes depending on the latent variable, we generate, for the same context drawn from the complexity level 3, several different outputs, only changing the latent variable realization.

\begin{figure}[!th]
\begin{center}
	\includegraphics[width=0.8\columnwidth]{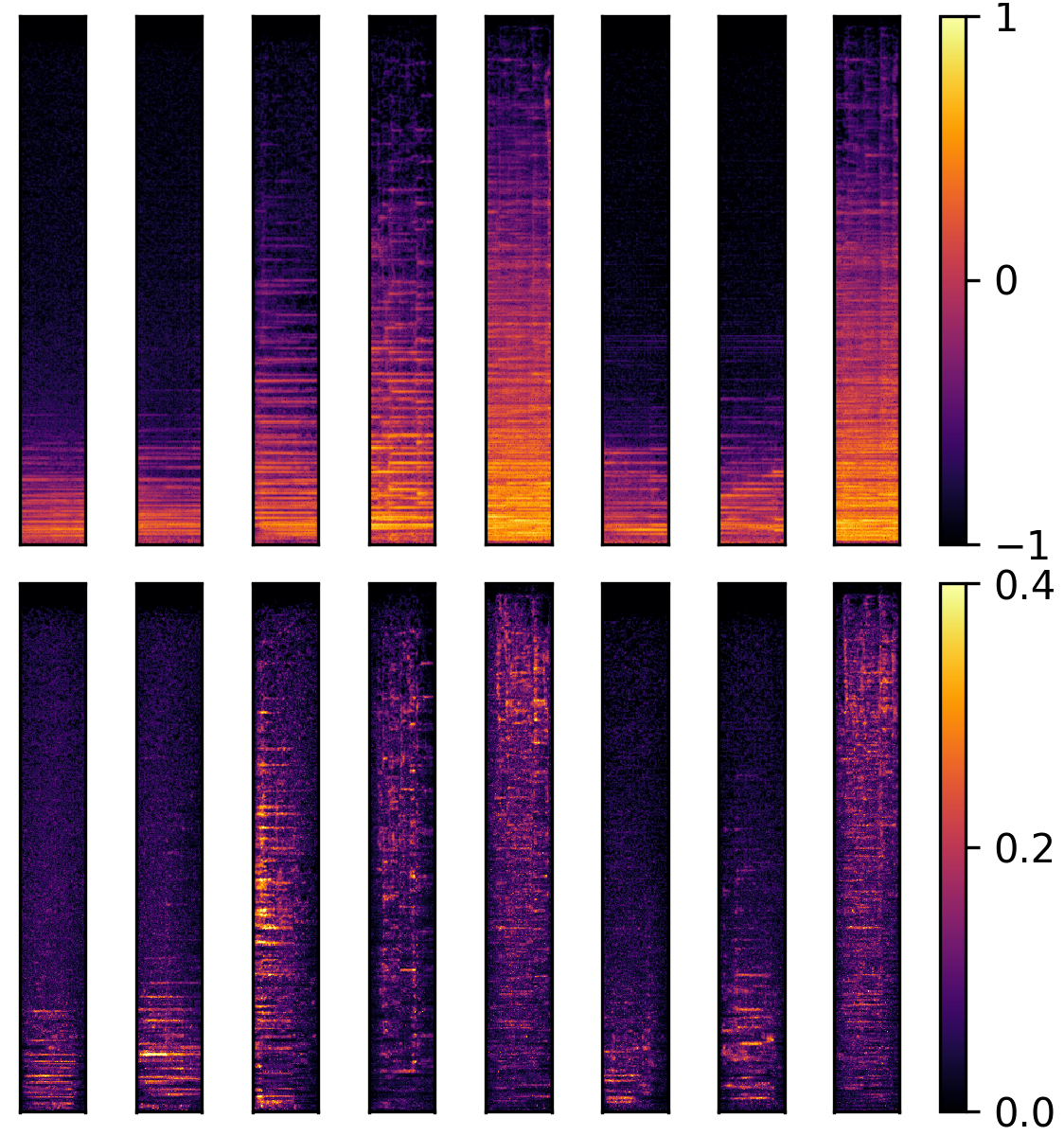}
	\caption{Top: Each column represent the mean of 8 generated gaps drawing different samples from the latent variable and keeping the context fixed. Bottom: Standard deviation of those 8 gaps.}
	\label{fig:noise}
\end{center}
\end{figure}

\begin{figure}[!th]
\begin{center}
	\includegraphics[width=0.9\columnwidth]{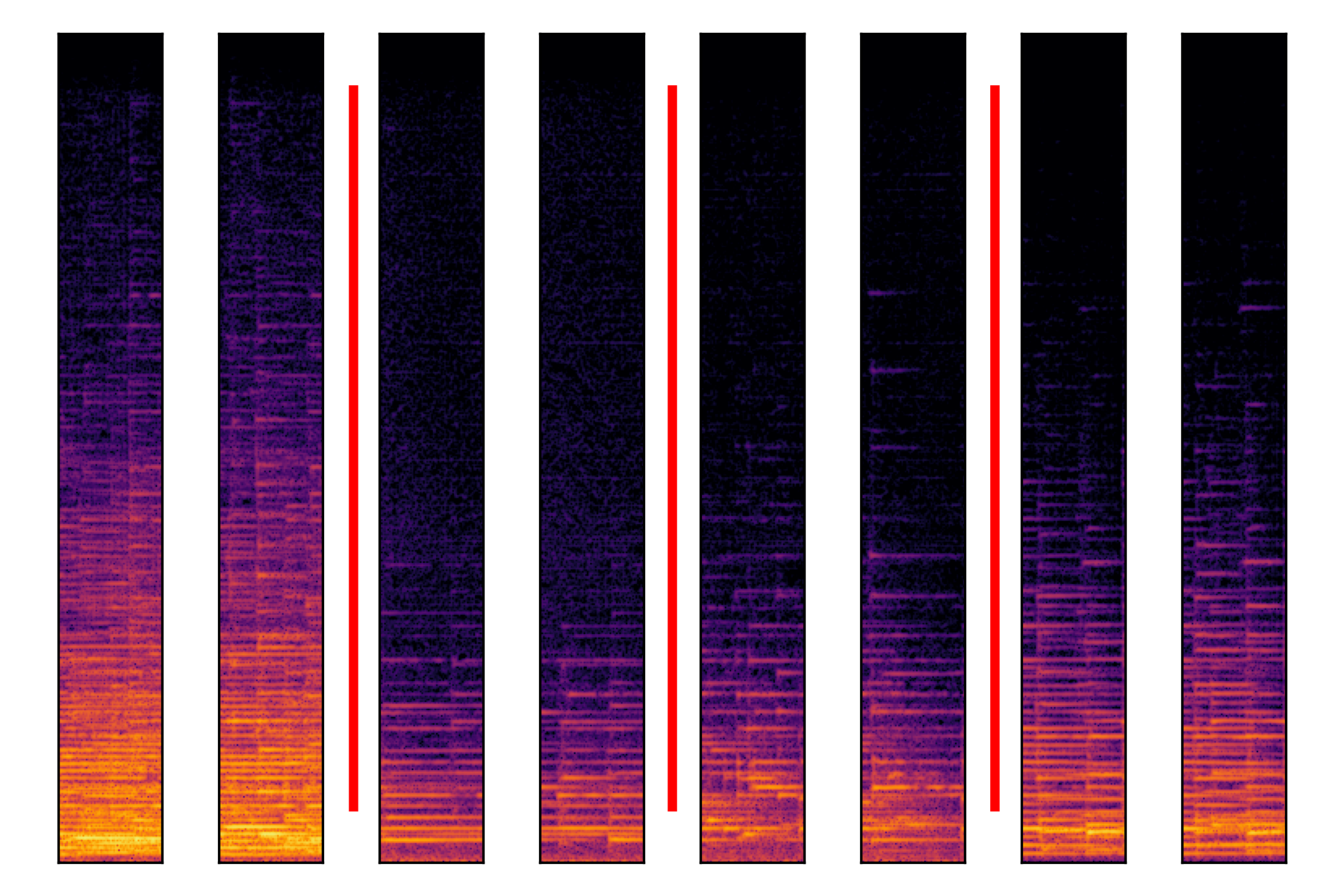}
	\caption{Every two columns separated with red lines show two different outputs of the system for the same contexts and different samples from the latent variable.}
	\label{fig:noise-2}
\end{center}
\end{figure}

The mean and standard deviation of the generated spectrograms for eight different samples from the latent variable and eight different contexts from the maestro dataset are shown in Fig. \ref{fig:noise}. We can see here that the mean is not completely blurred, but the standard deviation is not small. This indicates that the output does drastically change with the different samples from the latent variable, but there is still significant variation. Going into more singular cases, in Fig. \ref{fig:noise-2}, we can see 4 pairs of gaps generated with 4 contexts and different realizations of the latent variable. On these, the differences on the spectrograms are clear and exemplify what we observed by analyzing a larger batch of examples. Differences can manifest, e.g., changes to the intensity with which some notes are played or modified chord sequences. 
Additionally, in our website\footnote{\url{https://andimarafioti.github.io/GACELA/}} we provide sound samples for the examples from Fig. \ref{fig:noise-2}. These sound samples are clearly distinguishable from one another.

\subsection{Attention of the encoder}

Our system encodes the context of the lost information in order to use it as conditioning for a generative network. A key variable here is how much context is encoded by the system since the amount of context is proportional to the computation time. Therefore, we evaluate how the context is being exploited by the network; we provide the encoder with different contexts drawn from the complexity level 3 and analyse the output it produces. Since the encoder is comprised only of convolutional layers, and convolution is translation equivariant, the encoder outputs preserves the localization of the information. Hence by analysing the output, we know which part of the input mel-spectrogram was encoded for the network to decode a solution for the gap.

\begin{figure}[!th]
\begin{center}
	\includegraphics[width=0.8\columnwidth]{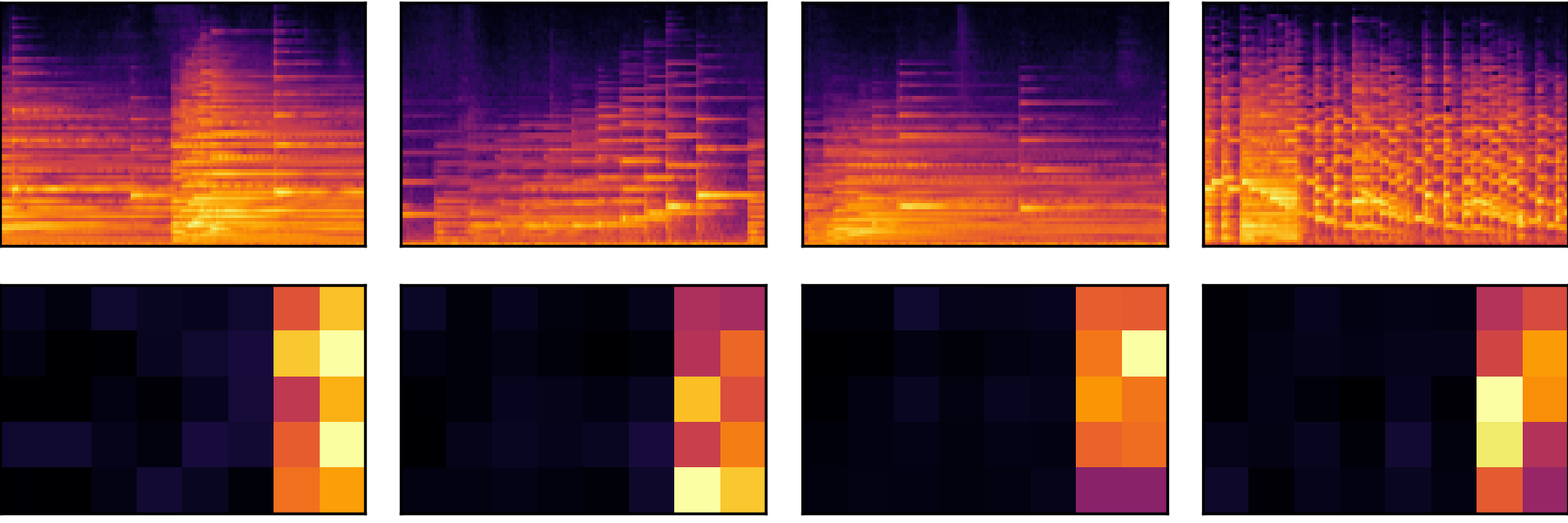}
	\includegraphics[width=0.8\columnwidth]{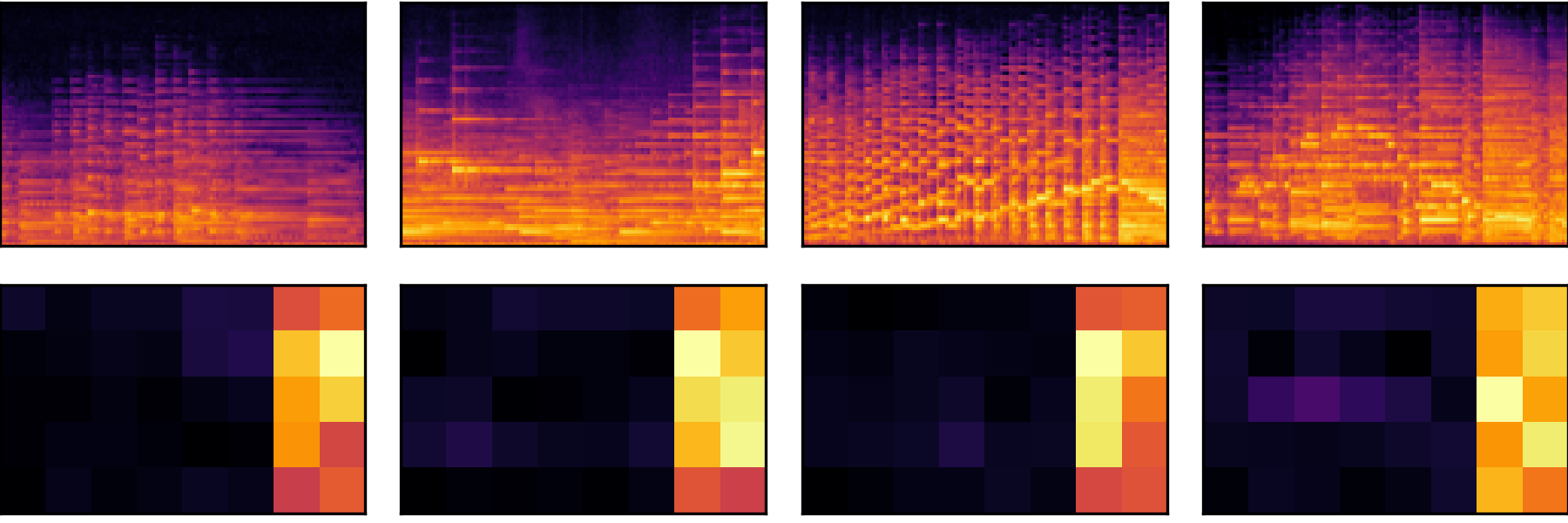}
	\caption{Two repeated rows where top is input spectrograms for the encoder (after time and mel average) and bottom is the encoder output averaged across channels.}
	\label{fig:attention}
\end{center}
\end{figure}

\begin{figure}[!th]
\begin{center}
	\includegraphics[width=0.8\columnwidth]{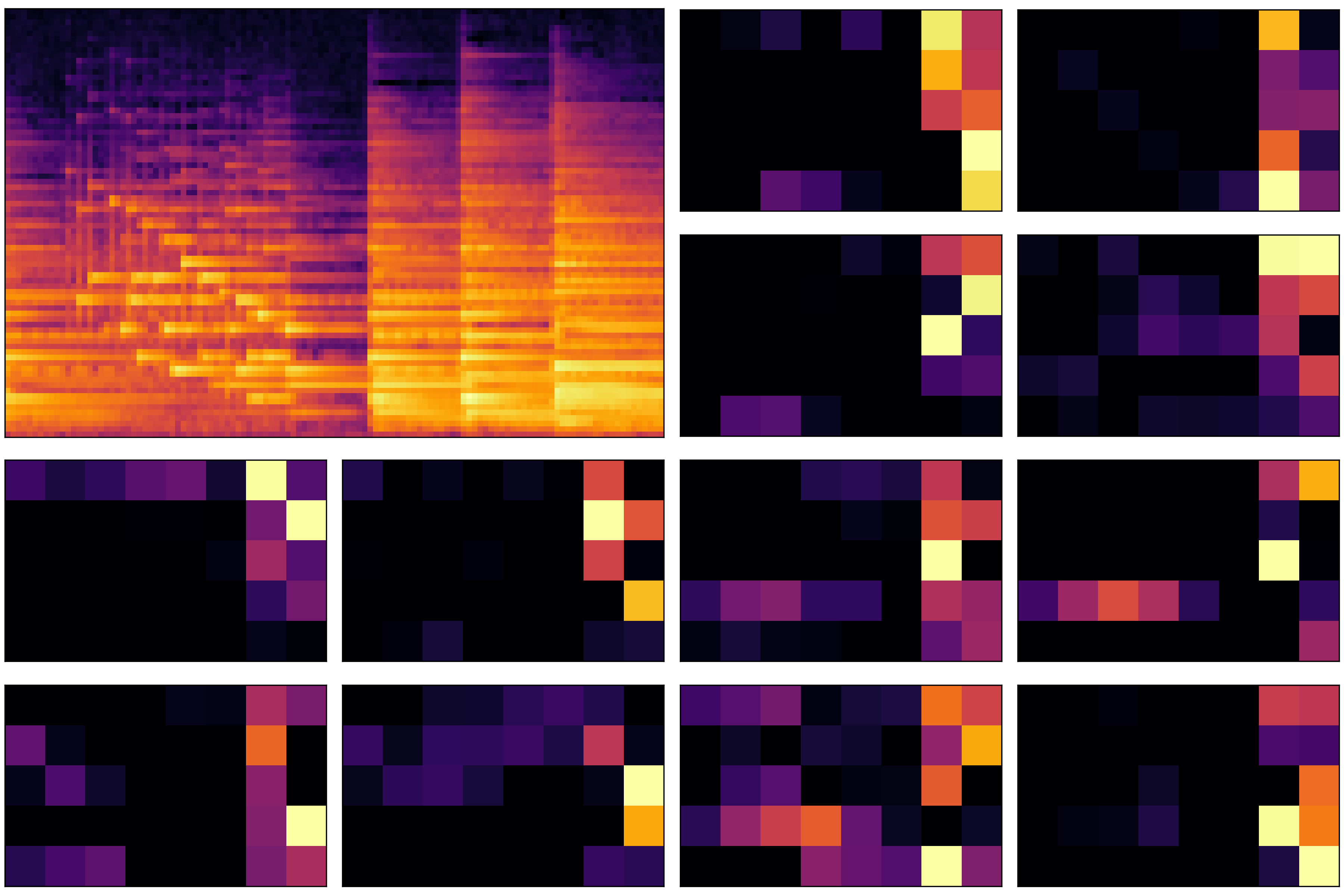}
	\caption{Top left is input spectrograms for the encoder (after time and mel average). Others are channels for the output of the encoder.}
	\label{fig:attention-2}
\end{center}
\end{figure}

Fig. \ref{fig:attention} shows different spectrograms that were given as input to the encoder before the gap and the average encoded output across channels. The sparse nature of the code tells us that the encoder puts his attention mostly on the two time bins adjacent to the gap. In the signal domain, this corresponds to roughly 1.4 seconds of audio content, when the full context represents 5.6 seconds. While not displayed, the situation is symmetric for the post-gap encoder.
Fig. \ref{fig:attention-2} shows different channels of the encoders output for one particular input. We observe here that even though the mean is focusing on the information closer to the gap, the rest of the mel-spectrogram is still useful and encoded.

\subsection{Comparison to the similarity graph algorithm.}

Even though both the similarity graph algorithm (SGA) and our method inpaint gaps of audio content in a similar range, they rely on very different conditions: 
a) SGA should have sufficient material to compute the similarity graph on and find a suitable solution, e.g., a full song. In contrast, the neural network only takes a few seconds of audio context.
b) SGA does not require any training and can adapt to multiple datasets and gap lengths, while our network needs hours of re-training for each new type of music and (currently) target gap length.
c) SGA may modify the length of the gap, as well as the total length of the song and some uncorrupted audio content at the border of the gap. Our method only replaces the corrupted portion of the song.
d) Once trained, filling the gap using the network is computationally more efficient than SGA.

Since the conditions for both methods are so different, they suffer from different drawbacks. SGA always fills the gap with content that has the same audio quality as the rest of the piece, since it fills the gap with content from the piece. However, the two transitions between the borders around the gap can be unnatural if the algorithm picks an unsuitable gap replacement. Furthermore, this selection relies crucially on the existence of a good replacement in the current song. Eventually, SGA results can be erratic with the reconstruction being either very good or relatively poor with a good/poor ratio depending highly on the specific piece of music it is applied to.
Our network is limited differently as it can only access a few seconds of information around the gap and only works on the type of data it has been trained on. Therefore, the perceived disturbances in the gap reconstruction are different from SGA. In general, the quality of the reconstruction is more uniform and the transitions are not the main source of artifacts.

Under these considerations, a comparison to SGA in terms of listening tests is outside of the scope of this contribution. There are conditions for which SGA provides an optimal solution: A localized corruption within a repetitive song, where the solution does not need to be computed quickly and a change in song duration is unproblematic. On the other side, our method can handle other conditions such as streamed signals where the full content is not available, signals that present repetitive corruptions in intervals such that long context is not available, or signals without repetition. Nevertheless, we applied our method trained on the complexity level 4 to the rock songs used for the listening tests in \cite{Perraudin2016} and provide them on the webpage\footnote{\url{https://andimarafioti.github.io/GACELA/}}.


\subsection{General perceptual impact: detection and severity.}

\begin{figure}[!th]
\begin{center}
	\includegraphics[width=\columnwidth]{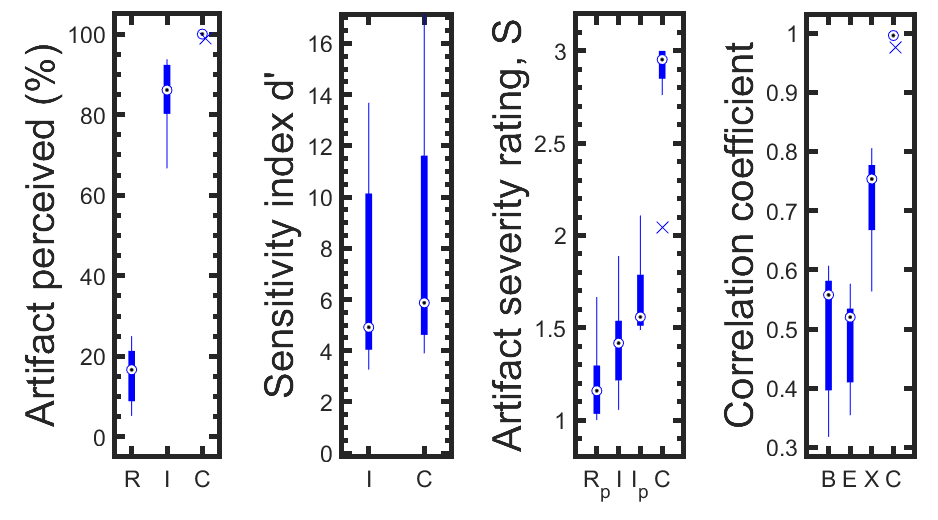}
	\caption{Perceived artifacts across all subjects. Left panel: Statistics of the detection rate. Left-center panel: Statistics of the sensitivity index $d'$, i.e., the artifact-detection rate relative to the false-alarm rate, with $d'=1$ corresponding to the chance rate. Right-center: Statistics of the severity ratings. Right: Statistics of the correlation coefficients between the perceived artifact position versus begin (B), end (E), and best choice of B and E (X) of the artifact in the inpainting condition. Conditions: reference ($R$), inpainted (I), and click (C), reference when perceived as artifact ($R_p$), inpainted when perceived as artifact ($I_p$). Statistics: Median (circle), 25\% and 75\% quartiles (thick lines), coverage of 99.3\% (thin lines, assuming normal distribution), outliers (crosses, horizontally jittered for a better visibility).}
	\label{fig:det-rate-general}
\end{center}
\end{figure}

Detection results are shown in the left panel of Fig. \ref{fig:det-rate-general}. The average detection rates for the click, inpainting, and reference conditions were
99.9$\pm$0.4\%, 84.7$\pm$9.3\%, and 15.6$\pm$7.3\%, respectively. The almost perfect detection rate and small variance in the click condition demonstrates a good attention of our subjects, for whom even a single click was clearly audible. The clearly non-zero rate in the reference condition shows that our subjects were highly motivated to find artifacts. The detection rate in the inpainted condition was between those from the reference and click conditions. Note that the reference condition did not contain any artifacts, thus, the artifact detection rate in that condition is here referred to as the false-alarm rate. The large variance of the false-alarm rate shows that it is listener-specific. Thus, for further analysis, the detection rates from the inpainted condition were related to the listener specific false-alarm rate, i.e., the sensitivity index d' was used \cite{macmillan2004detection}. The false-alarm rate can be considered as a reference for guessing, thus, d' = 1 indicates that the artifacts was detected at the level of chance rate. The left-center panel of Fig. \ref{fig:det-rate-general} shows the statistics of d' for the inpainting and the click conditions. For the click condition, the average across all subjects was 8.2 $\pm$ 5.2, again demonstrating a good detectability of the clicks. For the inpainting condition, the average d' was 6.9 $\pm$ 4.2, i.e., slightly below that of the click. A t-test performed on listeners' d' showed a significant (p = 0.018) difference from click-detection, indicating that our listeners, as a group, were less able to detect the inpainting than the click condition. 

The center-right panel of Fig. \ref{fig:det-rate-general} shows the statistics of the severity ratings reported in the real, inpainted and click conditions. For the click condition, the ratings were close to 3 ("not acceptable") with an average across all subjects of
2.83 $\pm$ 0.33. This indicates that on average, our subjects rated
the clicks as not acceptable. In contrast, for the inpainted condition,
the average rating was 1.41$\pm$0.26, between ``not disturbing'' and ``mildly disturbing''. This average considers undetected inpainted signals, rated with a 0. The average rating for detected inpainted signals was 1.66 $\pm$ 0.24. This is still significantly
($p<0.001$) lower than the severity of the clicks as revealed
by a paired t-test calculated between the ratings for clicks and
inpainted for detected artifacts. This indicates that when
the inpainting artifacts were perceived, their severity was rated
significantly lower than that of the clicks. Additionally, when the reference signals were classified as having an artifact, they were on average rated across all subjects with 1.21$\pm$0.22.

The right panel of Fig. \ref{fig:det-rate-general} shows the average correlation coefficients between the perceived position of the inpainting and its actual position. The correlations indicate that as soon as our subjects detected an artifact, they had some estimate of its position within the stimulus. For the clicks, the higher correlation indicates that our subjects were able to exactly determine and report the position of the click.

\subsection{Effect of the complexity level}

\begin{figure}[!th]
\begin{center}
	\includegraphics[width=\columnwidth]{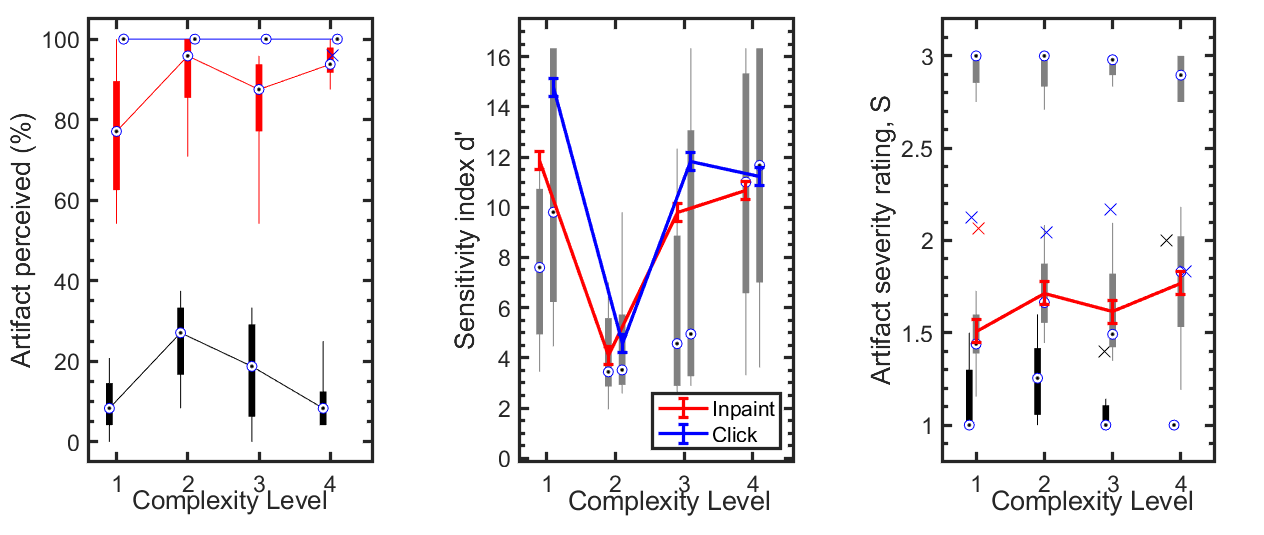}
	\caption{Effect of the complexity level. Left: Statistics of the detection rate. Center: Sensitivity representing ability to detect artifacts shown as statistics (as in Fig. \ref{fig:det-rate-general}) and averages $\pm 1$ standard errors resulting from a three-way ANOVA. Right: Ratings of artifact severity shown as statistics (black: real; grey: inpainted; blue: click) and averages $\pm 1$ standard errors resulting from a two-way ANOVA. Click: sounds distorted with a click. Inpaint: sounds distorted with a gap of 750~ms and then inpainted by the GAN.}
	\label{fig:det-rate-per-music}
\end{center}
\end{figure}

The left panel of Fig. ~\ref{fig:det-rate-per-music} shows the percentages of artifact perceived for every condition as a function of the complexity level. The click was perfectly detected on every complexity level, but the false-alarm rate varied across the complexities. The center panel of Fig.~\ref{fig:det-rate-per-music} shows the statistics of sensitivity to detect an artifact as a function of the complexity level. The sensitivity index is related to the false-alarm rate, as can be seen in the large variations on the click sensitivity. In order to perform an statistical analysis, we did a three-way ANOVA on the sensitivities with the factors subject, complexity level, and type of distortion (inpainted and click) and their two-level interactions. The main effect of distortion was significant ($p<0.001$) indicating that the inpainting significantly reduced the rate of perceived artifacts. 

The main effect of complexity was significant ($p<0.001$) and its interaction with the type of distortion was significant ($p=0.007$) as well. A multiple post-hoc comparison (Tukey-Kramer test) showed that for the complexity levels of one and two, the rates in the click conditions were significantly ($p<0.05$) higher and lower, respectively, when compared to those for levels of three and four. This indicates that our subjects had a lower false-alarm rate in the most simple, MIDI-based piano sounds aligned to a regular grid, but higher in the still simple but not so-regular grid-based and more natural MIDI-generated piano sounds. Given such a tiny change in the tested sound material, the origin of such a large change in the detection rates can be explained by having our experiment run into ceiling effects -- The obtained sensitivities were high and well above the chance rate ($d'=1'$). Thus, the observed effect of individual complexity levels might be more related to random fluctuations at a ceiling of well detectable events than to a systematic impact of a factor. Thus, while we conclude that inpainting generally reduced the detection rate, more insight can be gained from the analysis of severity ratings. 


The right panel of Fig.~\ref{fig:det-rate-per-music} shows the ratings of artifact severity when an artifact was detected. We have performed a two-way ANOVA on the severity ratings with the factors subject and complexity level\footnote{The amount of data did not allow us to include the interaction between the subject and complexity level in that test}. The effect of the level was significant ($p=0.038$) indicating that across all subjects, the severity of the artifacts increased with the complexity. Despite its significance, the effect was small with all average ratings for inpaintings detected as artifacts being between ``not disturbing'' and ``mildly disturbing''. 

In summary, the network performance changed with the complexity of the music. For the detection rates, the results were not conclusive on the effect of the complexity. Nevertheless, the artifact severity ratings did vary per complexity, being better at lower complexities and showing that for every level of complexity the subjects found the inpainting to be better than `mildly disturbing'.

\subsection{Effect of the gap duration.}

\begin{figure}[!th]
\begin{center}
	\includegraphics[width=\columnwidth]{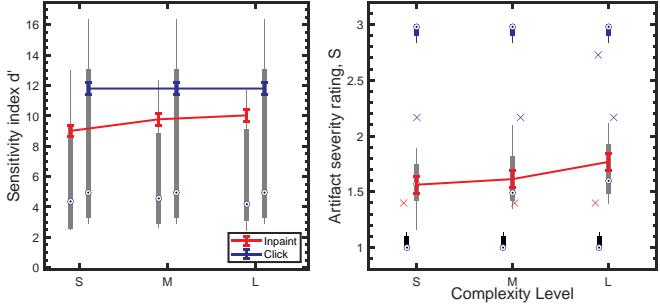}
	\caption{Effect of the gap duration. Left:  Sensitivity representing ability to detect artifacts shown as statistics (as in Fig. \ref{fig:det-rate-general}) and averages $\pm 1$ standard errors resulting from a three-way ANOVA. Right: Ratings of artifact severity shown as statistics (black: real; grey: inpainted; blue: click) and averages $\pm 1$ standard errors resulting from a two-way ANOVA. Click: sounds distorted with a click. Inpaint: sounds distorted with a gap and then inpainted by the GAN. S indicates a short gap of 375~ms, M a medium gap of 750~ms and L a long gap of 1500~ms.}
	\label{fig:det-rate-duration}
\end{center}
\end{figure}

The left panel of Fig.~\ref{fig:det-rate-duration} shows the statistics of sensitivity to detect an artifact as a function of the length of the gap on the third level of complexity (piano recordings). In order to perform an statistical analysis, we have performed a three-way ANOVA on the sensitivities with the factors subject, length of the gap, and type of distortion (inpainted and click) and their two-level interactions. The main effect of distortion was significant ($p<0.001$) indicating that the inpainting significantly reduced the rate of perceived artifacts. But in this case, the interaction between the distortion and the length of the gap was not significant ($p=0.4$), indicating that the improvement did not depend on the length of the gap. This is shown in that panel by averages and $\pm 1$ standard errors as a function of the complexity for the two types of distortion.

The right panel of Fig.~\ref{fig:det-rate-duration} shows the ratings of artifact severity when an artifact is detected. Their average values on the short inpainting (375~ms), medium inpainting (750~ms) and long inpainting (1500~ms) were 1.56$\pm$0.24, 1.61$\pm$0.27, and 1.77$\pm$0.45, respectively. This indicates that on average, our subjects rated the inpainting results between "not disturbing" and "mildly disturbing", even for the longer gaps of 1500~ms. We performed a two-way ANOVA on the severity ratings with the factors subject and length of the gap. The effect of the length of the gap was not significant ($p=0.17$), indicating that the ratings did not change with the gap length. When performing the same analysis on the full rating of the gaps (including non-detected gaps, rated as 0), the effect remained not significant ($p=0.07$).

In summary, the network performance did not change significantly for gaps between 375~ms and 1500~ms.

\section{Conclusions and outlook}

We introduced GACELA, a system for the restoration of localized audio information in gaps with a duration ranging between hundreds of milliseconds and seconds. GACELA is based on a conditional GAN and represents a further development based off our previous context encoder designed for audio inpainting of gaps up to tens of milliseconds~\cite{marafioti2019context}. The improvements consider two aspects. First, GACELA handles various time scales of audio information by considering five parallel discriminators with increasing resolution of receptive fields to prompt the generator's output to consider these time scales. Second, GACELA incorporates the inherent multi-modality of audio inpainting at this gap-duration range by being conditioned not only on the available information surrounding the gap but also on the latent variable of the conditional GAN. This provides the user with the option to fill-in the gap depending on his/her needs.

GACELA was evaluated numerically and in listening tests\footnote{We also encourage the interested reader to listen to the samples provided in \url{https://andimarafioti.github.io/GACELA/} to get a subjective impression on GACELA's performance.}. While under laboratory conditions our subjects were able to detect most of the inpaintings, the artifact severity was rated between ``not disturbing'' and ``mildly disturbing''. The detection rate and the severity ratings depended on the complexity of the sounds defined by the method of audio generation (MIDI vs. recordings) and number of instruments. The inpainted segments were more likely to be detected in sounds with larger complexity, with an exception found for the simplest complexity level represented as MIDI-generated piano music generated from artificial MIDI scores. Interestingly, our subjects were most sensitive to any type of corruption applied within this complexity level, confounding this part of results. The inpainting quality of GACELA did not change significantly for inpainting gaps with a duration ranging between 350~ms and 1500~ms. While it seems like GACELA could be applied to even longer gaps, we assume that the inpainting quality will drastically change at some gap duration. Also, as the training time increases with the gap and context durations -- GACELA might require improvements in order to be able to deal with significantly longer gaps. 

The results of our evaluation show the urgency of lowering the artifact-detection rate in the future. This is a very ambitious goal for long audio inpainting. The goal is to design a system receiving a new piece of music and generating a sound that is so similar that an attentive listener is not be able detect any artifacts. To this end, we expect future systems to better exploit auditory features and musical structures. In GACELA, the auditory features are represented by the compressed mel-spectrograms (conditioning the generator and input for three discriminators). In the future, generators directly producing features of an auditory space might provide improvements. A promising avenue in this regard are Audlet frames, i.e., invertible TF systems adapted to perceptually-relevant frequency scales~\cite{necciari18}. GACELA aims to preserve the musical structure by relying on discriminators handling various temporal scales separately. More explicit features such as beat- and chord-tracking, incorporated to both the training of neural networks as well as their assessment might further improve the inpainting quality in future systems.

\bibliographystyle{IEEEtran}
\bibliography{context}

\end{document}